\documentclass[12pt,preprint]{aastex}
\usepackage{natbib,graphicx}

\def\object#1{#1}
\def\rxj{RX J1605.3+3249}
\def\Sref#1{\S\,\ref{sec:#1}}
\def\Fref#1{Fig.~\ref{fig:#1}}
\def\Tref#1{Table~\ref{tab:#1}}
\def\Eref#1{Eq.~\ref{eq:#1}}
\let\la\lesssim\let\ga\gtrsim

\epsscale{0.55}

\makeatletter
\def\fps@figure{tbp}%
\def\fps@table{tbp}%
\def\thebib@list{%
 \list{\null}{%
  \leftmargin 3em\labelwidth\z@\labelsep\z@\parsep\z@\itemsep\z@
  \itemindent-\leftmargin\usecounter{enumi}%
 }%
}%
\makeatother

\begin{document}

\title{A Strong, Broad Absorption Feature in the X-ray Spectrum of the
Nearby Neutron Star RX~J1605.3+3249} 
\shorttitle{An Absorption Feature in RX J1605.3+3249}

\author{M. H. van Kerkwijk\altaffilmark{1},
D. L. Kaplan\altaffilmark{2},
M. Durant\altaffilmark{1},
S. R. Kulkarni\altaffilmark{2} and
F. Paerels\altaffilmark{3}}
\altaffiltext{1}{Department of Astronomy and Astrophysics, University of
Toronto, 60 St George Street, Toronto, ON~~M5S~3H8, Canada;
\texttt{mhvk,durant@astro.utoronto.ca}} 
\altaffiltext{2}{Department of Astronomy, California Institute of
  Technology, m.s.\ 105-24, Pasadena, CA~91125;
  \texttt{dlk,srk@astro.caltech.edu}} 
\altaffiltext{3}{Columbia Astrophysics Laboratory and Department of
  Astronomy, Columbia University, 538 W. 120th Street, New York,
  NY~10027; \texttt{frits@astro.columbia.edu}} 

\begin{abstract}
We present X-ray spectra taken with {\em XMM-Newton} of \rxj, the
third brightest in the class of nearby, thermally emitting neutron
stars.  In contrast to what is the case for the brightest object, RX
J1856.5$-$3754, we find that the spectrum of \rxj\ cannot be described
well by a pure black body, but shows a broad absorption feature at
27\,\AA\ (0.45\,keV).  With this, it joins the handful of isolated
neutron stars for which spectral features arising from the surface
have been detected.  We discuss possible mechanisms that might lead to
the features, as well as the overall optical to X-ray spectral energy
distribution, and compare the spectrum with what is observed for the
other nearby, thermally emitting neutron stars.  We conclude that we
may be observing absorption due to the proton cyclotron line, as was
suggested for the other sources, but weakened due to the strong-field
quantum electrodynamics effect of vacuum resonance mode conversion.
\end{abstract}


\section{Introduction}\label{sec:introduction}

The nearby, thermally emitting, radio-quiet neutron stars offer the
best possibilities for measuring spectra directly from a neutron-star
surface, uncontaminated by emission due to accretion and/or
magnetospheric processes.  Since the serendipitous discovery of the
first of these in \citeyear{wwn96} by Walter et al., six (possibly
seven) more have been uncovered in the ROSAT All-Sky Survey (see
reviews by \citealt{ttz+00,hab04}).  For four sources, optical
counterparts have been identified
(\citealt{wm97,mh98,kvk98,kkvk02,kkvk03}).  The high X-ray to optical
flux ratios leave no model but an isolated neutron star.  

At present, it is not clear what is the source of the thermal
emission.  The possibilities considered range from slow accretion from
the interstellar medium to release of residual heat, to decay of
strong magnetic fields.  Most important for the present purposes,
however, is that all six sources appear to have X-ray spectra that, as
far as one can tell from current observations, are entirely thermal,
thus offering the hope that it will be possible to do a `standard'
model-atmosphere analysis, and infer precise values of the
temperature, surface gravity, gravitational redshift and magnetic
field strength.

Given their interest, the nearby, thermally emitting neutron stars
have been among the prime targets for spectroscopy with the {\em
Chandra X-ray Observatory\/} and {\em XMM-Newton}.  First {\em XMM}
results on the second-brightest in the class, \object{RX
J0720.4$-$3125} (\citealt{pmm+01}), were somewhat disappointing, as no
lines were found.  The first {\em Chandra} spectra, of the brightest
thermally emitting neutron star, \object{RX~J1856.5$-$3754}, also
revealed no lines (\citealt{bzn+01}).  Indeed, even a much longer
(500~ks) integration failed to show evidence for any features, showing
instead a spectrum remarkably well described by a black body affected
only by interstellar extinction (\citealt{dmd+02,br02,bhn+03}).
Similarly, {\em Chandra} data on the \object{Vela pulsar}
(\citealt{pzs+01}) and \object{PSR B0656+14} (\citealt{ms02}), both of
which have strong thermal components in their X-ray spectra, failed to
show any features.

The only exceptions came recently.  First, \citet{spzt02} discovered
two absorption features in {\em Chandra} observations of 1E
1207.4$-$5209, the pulsating central source in the supernova remnant
PKS 1209$-$51/52.  Variability in these features as a function of
pulse phase was seen in {\em XMM} observations by \citet{mdlc+02}, and
confirmed by \citet{bclm03}, who also reported a third and possibly
even a fourth feature, all harmonically spaced.  The nature of the
absorption features is not yet clear, with suggestions ranging from
cyclotron lines to transitions of Helium or mid-Z atoms in a strong
magnetic field (\citealt{spzt02,hm02}).  Second, \citet{hsh+03}
discovered a broad absorption feature in {\em XMM} spectra of RX
J1308.6+2127, a nearby, thermally emitting neutron star.  As before,
it is not clear what causes the feature; \citeauthor{hsh+03} speculate
it might be due to proton cyclotron absorption.  Third, while we were
revising of our manuscript, a preprint by \citet{hztb03} showed that
another nearby neutron star, RX J0720.4$-$3125 had a broad, but weaker
absorption feature as well, contrary to earlier claims
(\citealt{pmm+01,pzs+01,kvkm+03}). 

Here, we present the discovery of a broad absorption feature in a
third nearby, thermally emitting neutron star, \rxj.  We describe our
observations and reduction in \Sref{data}.  In \Sref{spectrum}, we
analyze and characterize the spectrum, and in \Sref{timing}, we derive
limits to any periodic variations.  We discuss implications for our
understanding of \rxj, as well as the nearby, thermally emitting
neutron stars in general, in \Sref{speculation}.

\begin{deluxetable}{lllllll}
\tablecaption{Summary of X-ray Observations\label{tab:log}}
\tabletypesize{\small}
\tablewidth{0pt}
\tablehead{
\colhead{ObsID}& \colhead{Instrument}& \colhead{Mode}& \colhead{Start Time}&
\multicolumn{2}{c}{Exposure}& \colhead{Count Rate\tablenotemark{a}}\\
& & & &\colhead{Raw} &\colhead{Filtered} & \\
& & &\colhead{(UT)} &\colhead{(ks)} &\colhead{(ks)}&
\colhead{(${\rm s}^{-1}$)}}
\startdata
2791 & ACIS-I & Standard & 2002~Jan~07.17 & 20.4 & 20.4 & 0.153\tablenotemark{b}\\[0.1in]
0073140301 & EPIC-PN & Timing &       2002~Jan~10.04 & 26.0 &14.3 &3.118(16)\\
 & RGS1/RGS2 & Standard & & 33.6 & 20.9/19.4 & 0.139(3)/0.130(3)\\[0.1in]
0073140201 & EPIC-PN & Timing &       2002~Jan~15.99 & 26.4 &24.2 &3.083(12)\\
 & RGS1/RGS2 & Standard & & 30.4 & 27.8/27.0 & 0.138(3)/0.127(3)\\[0.1in]
0073140501 & EPIC-PN & Timing &       2002~Jan~19.97 & 29.8 &18.4 &3.069(14)\\
 & RGS1/RGS2 & Standard & & 33.9 & 22.2/21.5 & 0.141(3)/0.133(3)\\[0.1in]
0157360401 & EPIC-PN & Large-window & 2003~Jan~17.91 & 33.2 &25.7 &2.385(10)\\
 & RGS1/RGS2 & Standard & & 41.7 & 31.3/31.2 & 0.144(2)/0.133(2)\\[0.1in]
0157360601 & EPIC-PN\tablenotemark{c} & 
                       Large-window & 2003~Feb~26.80 & 16.8 & 8.8 &1.477(13)\\
 & RGS1/RGS2 & Standard & & 32.2 & 10.1/\phn9.5& 0.148(5)/0.147(6)\\
\enddata
\tablenotetext{a}{All count-rates are background subtracted and only from
the filtered time intervals.  Count-rates for EPIC-pn are for single
events with energies $>0.3$~keV.  Numbers in parentheses indicate
uncertainties in the last digit.}
\tablenotetext{b}{Piled-up.}
\tablenotetext{c}{Taken with the thick filter.  All other EPIC
  observations are taken with the thin filter.}
\end{deluxetable}

\section{X-ray observations}\label{sec:data}

We observed \rxj\ with {\em XMM} three times, on 9, 15, and 19 January
2002, for a total of approximately 100\,ks.  In addition, we analysed
{\em XMM} data taken for calibration purposes on 18 January and 27
February 2003, and data taken with {\em Chandra} on 7 January 2002.  A
summary is given in Table~\ref{tab:log}.

\subsection{XMM}

{\em XMM} consists of three X-ray telescopes (\citealt{jla+01}).  For
two of these, half the light is deflected to Reflection Grating
Spectrometers (RGS; \citealt{dhbk+01}), while the other half is fed to
European Photon Imaging Cameras with MOS detectors (EPIC-MOS;
\citealt{taa+01}).  A third camera, with PN detectors (EPIC-PN;
\citealt{sbd+01}) receives all the light of the third telescope.

\subsubsection{EPIC-MOS imaging}

As \rxj\ is bright, photon pile-up, where multiple photons arrive in
one integration time, is a concern.  We had hoped to use the EPIC MOS
detectors in a mode in which the central chip is read out fast, but
this had not yet been commissioned at the time of the observations.
We decided to use full-frame mode instead (2.6-s frame time),
sacrificing spectral and timing information for field of view, hoping
to increase the number of background sources and hence improve the
absolute astrometry.  This objective became moot with the measurement
of an accurate {\em Chandra} position: $\alpha_{\rm J2000} = 16^{\rm
h}05^{\rm m}18\fs50\pm0\fs06$, $\delta_{\rm J2000} = +32^\circ
49'17\farcs4\pm0\farcs7$ (\citealt{kkvk03}).  The observations were
taken through the thin filter, in order to maximize the soft response.
The same setup happened to be used for the calibration observations.

We reprocessed the three observations using the standard task {\sc
emchain} in the {\em XMM} Science Analysis System (XMM-SAS), version
5.4.1.  We determined the positions of \rxj\ in both cameras and all
observations and found these to be fully consistent with the {\em
Chandra} position mentioned above (within 3\arcsec\ before aspect
correction).  We used the MOS positions in the RGS and PN-timing
reduction to define the source position in the local, satellite frame.

\begin{figure}
\plotone{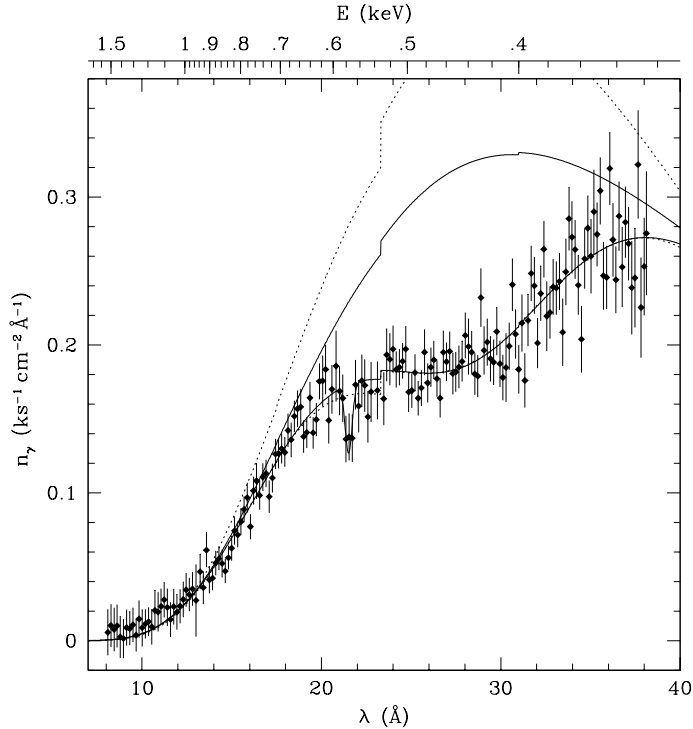}
\caption[]{Flux-calibrated RGS spectrum of RX J1605.3+3249, binned to
  0.175\,\AA.  At this binning, all points are uncorrelated,
  independent flux measurements.  The total integration time is
  110\,ks.\label{fig:rgs}
  Overdrawn are two fits: slightly extincted black body
  curves with one (lower dotted line) and two Gaussian (lower
  continuous curve) absorption features.  The upper curves reflect the
  corresponding continua ($n_{\rm c}(\lambda)$ in
  \Eref{gauss}).\label{fig:rgsmodel}} 
\end{figure}

\subsubsection{RGS spectroscopy}

For the RGS observations, we used the standard `High Event Rate with
SES' spectroscopy mode for read-out.  We reprocessed the data using
the XMM-SAS routine, {\sc rgsproc}.  We first made event files and
determined times of low background from the count rate on CCD~9 (which
is closest to the optical axis), rejecting all times that the rate was
above 0.5$\rm\,s^{-1}$.  The final exposure times and net count rates
are listed in Table~\ref{tab:log}.  Source and background counts were
extracted using the standard spatial and energy filters; for the
source position, which defines the spatial extraction regions as well
as the wavelength zero point, we used the position inferred from the
corresponding EPIC-MOS camera and exposure.  As the source showed no
sign of variability, we combined all spectra into one flux-calibrated
average using the task {\sc rgsfluxer}.  The result, binned to
0.175\,\AA, is shown in \Fref{rgs}.

\subsubsection{EPIC-PN spectroscopy}
\label{sec:obs:xmm:pn}

\paragraph{Timing mode.}  For the 2002 PN observations, we used the
timing mode, in order to avoid pile-up and to allow a search for
variability at as large a range of periods as possible.  In this mode,
the central CCD is read out continuously, and hence the gain in time
resolution comes at the cost of loss of positional information along
the detector columns and increased background.  The observations were
taken through the thin filter in order to maximize the response at low
energies.

We reprocessed the data using the XMM-SAS task {\sc epchain}, using
the EPIC-MOS1 position as a reference for timing purposes.  We
extracted source spectra in a region of 17 pixels centered on the peak
of the (one-dimensional) point-spread function ($31\leq{\rm
rawx}\leq47$).  For the background, we used a region of the same size
away from the source ($6\leq{\rm rawx}\leq22$).  For both, we excluded
times of high background due to proton flares (rejecting all 52-s
intervals in which the 0.3--1\,keV background count rate exceeded
0.25\,s$^{-1}$).

\begin{figure}
\plotone{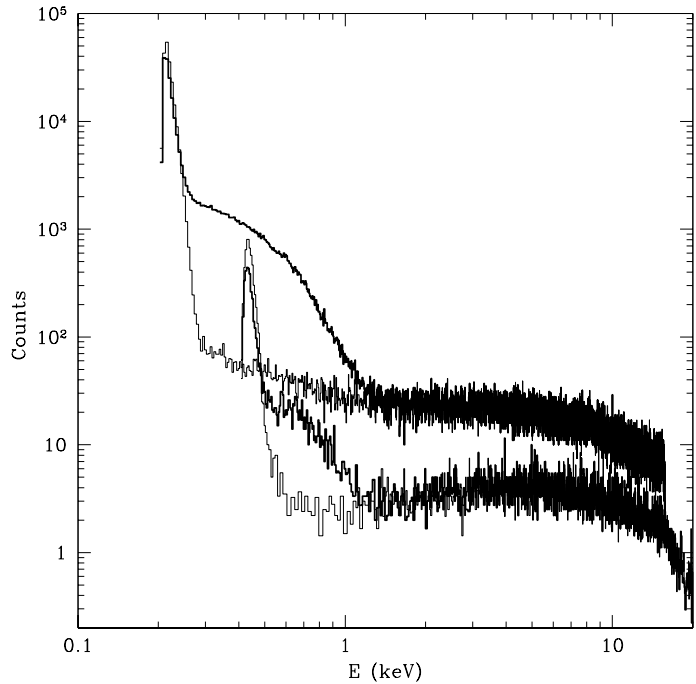}
\caption[]{Count spectra for the timing mode EPIC-PN data taken on 15
January 2002.  The top two curves are for the single-pixel events for
the source (bold) and background region, while the bottom two curves
are for the double-pixel events.  The strong peaks at low energies are
instrumental, and have different strength for the source and
background regions.  If not taken into account, the peak in the
double-pixel events will lead to artificial features in the
background-corrected spectrum.\label{fig:sd}}
\end{figure}

Looking in detail at the count spectrum for the timing data (see
\Fref{sd}), we found a peculiar excess of counts with energies just
above 0.2\,keV for single-pixel events, and energies just above
0.4\,keV for double-pixel events.\footnote{In single-pixel events, all
of the photon energy has been absorbed in a single pixel.  If a photon
arrives closely to the border of a CCD pixel, the energy will be
spread over more pixels.}  Inspecting the event file, it seems that
these are due to flickering, `hot' pixels, which lead to bursts of
events over short periods of time.  Many of these follow each other
sufficiently rapidly that in the data stream they appear as
neighbouring events and thus are identified by the pipeline as
double-pixel events.  These hot pixels pose a problem, as their number
is not the same in the source and background regions.  Hence, if not
taken into account, one will obtain instrumental emission or
absorption features around 0.2 and 0.4\,keV.  The latter is exactly in
the range where we find absorption in the RGS spectra
(\Sref{spectrum:rgs}).

Because of the above, for our spectroscopic analysis we had at first
decided to use only the single-pixel events, generating response
matrices for this selection using the appropriate SAS tasks.  However,
we learned (F.\ Haberl, 2003, personal comm.) that for single-pixel
events in timing mode the response is not very well understood, and
that only the combination of singles and doubles at energies above
0.5\,keV gives reliable, reproduceable results.  Indeed, the SAS task
generates a response valid for this combination (R.\ Saxton, 2003,
personal comm.; oddly, the results were consistent nevertheless -- see
\Sref{pnspec}).  It seems possible to circumvent these problems by
careful excision of the burst of noise events (\citealt{bhf+03}), but
we decided to forego this excercise, since for the large-window data
the response at low energies is fairly secure.

\paragraph{Large window mode.}  For the calibration data taken in
large-window mode, we again ran the standard task {\sc epchain}, and
rejected time intervals with strong background flaring using the
standard good-time intervals (for which the all-chip 7--15\,keV count
rate is below 10\,s$^{-1}$).  We selected source events within an
aperture of $64\farcs5$ radius, and background events from an annulus
with radii between 80 and $100\arcsec$; this means that our background
is not all taken from the same chip, but given the very large source
count rate, the effect of this is minimal.  Indeed, the count rate of
the source is sufficiently high, $2.4\rm\,s^{-1}$, that despite the
48-ms frame time offered by large-window mode there will be some pile
up: in some integrations two photons will hit the detector so closely
to each other that the resulting charge distribution will be
identified as a single event (with anomalously high energy).  Our
source is very soft, essentially following a Wien tail at higher
energies; as a result, even a small fraction of piled up events can
give the false impression of a hard tail.  Comparing the spectra of
single and double-pixel events, we find this to be the case.
Therefore, we decided to use single-pixel events only.\footnote{As the
{\em XMM} point spread function is well-resolved, the probability of
two photons being absorbed fully in a single pixel during one
integration time (leading to a single-pixel event with overestimated
energy) is much smaller than the probability of two photons being
fully absorbed within two neighbouring pixels or at least one being
spread over two pixels (leading to a double-pixel event).}  We note
that even with those, we see some evidence for pile up at the higher
energies.  As it is a small effect, however, we felt it not worthwhile
to try to ameliorate this by cutting out the core of the point-spread
function, in particular as the uncertainty in how the point-spread
function varies as a function of energy may well be larger than any
residual effects due to pile-up.  Instead, we simply conclude that
while the large-window mode data should give a more secure view of
what happens at low energies, the timing-mode data are to be trusted
better at high energies.

\subsection{Chandra}

For the {\em Chandra} data, we first reprocessed the level-1 event
file into a level-2 event file, following standard
procedures,
while keeping the events that had been flagged as possible
afterglow.\footnote{See
\url{http://asc.harvard.edu/ciao/threads/acisdetectafterglow/}.}  We
also took advantage of an updated response file and corrected for the
charge transfer inefficiency of the ACIS S3 detector, on which the
source was located.
We used the \texttt{CIAO} tool {\sc psextract} to extract source and
background spectra -- from a circular region centered on \rxj\ with a
radius of 4~pixels, and an annulus from radii of 20~pixels to
45~pixels, respectively -- as well as to create the response files.  To
account for the slight degradation with time of the ACIS detectors, we
applied a correction appropriate for our observing date.\footnote{See
\url{http://asc.harvard.edu/ciao/threads/apply\_acisabs/}.}  Due to
the 3.2-s frame time of the ACIS-I observations, \rxj\ suffers from
significant pileup (the count-rate was $\approx 0.15\mbox{ s}^{-1}$).
We therefore used the \texttt{Sherpa} pileup model
\citep{dav01} when analyzing the data.  We note that the Sherpa
pile-up model can give erroneous results if one tries to determine
which spectral shape fits the data best.  It should suffice, however,
for our purpose of cross-checking the {\em XMM} results.

\section{The spectrum}\label{sec:spectrum}

The flux-calibrated RGS spectrum (\Fref{rgs}) is clearly inconsistent
with a smooth function like an absorbed black body, unlike what is
seen for RX J1856.5$-$3754, the brightest thermally emitting neutron
star (see \Sref{introduction}).  In order to quantify the departures,
we tried fitting black bodies with simple absorption features, and
compare the results between the different instruments.

Before describing the results, we should stress that there are still
energy-dependent inconsistencies between the calibration of EPIC-pn
and RGS at the $\la\!20$\% level (M.~Sako, J.-W. den Herder, 2003,
personal communications; \citealt{kir03}; \Fref{unfolded} below),
especially below $\sim\!0.5\,$keV, and it is not known which is
correct.

\subsection{Fitting the RGS Data}\label{sec:spectrum:rgs}

We start with the RGS data, as these have the best resolution.  We
rebin them to 0.175\,\AA, as shown in \Fref{rgs}.  As expected from
the figure, a black body gives an unsatisfactory fit, with a reduced
$\chi^2$ of 2.6 (for 170 degrees of freedom), and with a best-fit
column density $N_H$ of zero (see \Tref{fits}).

As a first-order parametrisation of the deviations from a black body,
we include Gaussian absorption features, as follows,
\begin{equation}
n(\lambda)=n_{\rm c}(\lambda) \prod_i\left[1-r_{i}\exp \left(
  -4\log 2\frac{(\lambda-\lambda_{i})^{2}}{{\rm FWHM}_i^{2}}\right)\right]
\label{eq:gauss}
\end{equation}
where $n(\lambda)$ is the photon rate per unit wavelength at
wavelength $\lambda$, $n_{\rm c}(\lambda)$ the continuum photon rate
(in our case the extincted black body), and $r_i$, $\lambda_i$ and
FWHM$_i$ are the fractional depth, central wavelength, and full width
at half maximum of the Gaussian feature.

We find that the inclusion of a single Gaussian improves the quality
of the fit dramatically, changing the reduced $\chi^2$ from 2.6 to
0.9, i.e., a good fit (see \Fref{rgsmodel}; \Tref{fits}).  The
Gaussian is centered at 27.3\,\AA\ (0.45\,keV), and has a FWHM of
13\,\AA\ (0.2\,keV).  Since this Gaussian is so broad, covering a
significant fraction of the spectrum, its amplitude is highly
covariant with the other fit parameters that determine the overall
shape of the spectrum ($N_{\rm H}$, $kT$, and $R_\infty/d$); all
these, therefore, have rather large uncertainties.

In \Fref{rgsmodel}, it looks like there is also a narrower absorption
feature, at $\sim\!22\,$\AA.  We included a second Gaussian in the fit
and found a central wavelength of 21.5\,\AA\ (0.58\,keV), and a FWHM
of 0.5\,\AA\ (0.12\,keV); see \Tref{fits}.  This wavelength is close
to that of the resonance line of He-like Oxygen (\ion{O}{7}), but it
would be hard to understand as interstellar absorption, given both the
large equivalent width and the fact that it is resolved.  It is
unlikely to be an instrumental artifact, as we do not see it in RGS
spectra of other thermally emitting neutron stars.  Thus, it may come
from the neutron-star surface.  We stress, however, that the detection
needs to be confirmed: while the line itself is significant at the
3.5-$\sigma$ level, the significance of the feature decreases to
marginal once one takes into account properly the number of trials (of
order 100) associated with looking anywhere in the spectrum for
absorption features.  Independent of its significance, one thing to
note is how strongly the addition of even this small line affects the
inferred continuum (see \Fref{rgsmodel}).  This is due to the
above-mentioned covariance between the various parameters.

\begin{deluxetable}{lrrrrrr}
\tablewidth{0pt}
\tablecaption{Fits to the RGS and EPIC spectra\label{tab:fits}}
\tablehead{
\colhead{Parameter} & 
\multicolumn{3}{c}{\dotfill RGS\dotfill}&
\multicolumn{3}{c}{\dotfill EPIC\dotfill} \\
 & \colhead{0-G} & \colhead{1-G}& \colhead{2-G} &
   \colhead{0-G} & \colhead{1-G$_E$}& \colhead{1-G}}
\startdata
$N_{H}$ $(\times 10^{20}{\rm cm}^{-2})$ \dotfill
                                        & 0 & 2.2(11) & 0.8(10) & 
                                          0 & 0.68(10)& 0.98(19)\\[0.1in]
$kT$ (eV) \dotfill &              105.0(11) & 90(4)  & 95(4)   &  
                                   98.00(15)& 94.1(5)& 92.6(8)\\
$R_{\infty}/d$ (km/kpc) \dotfill &  6.66(17)& 15(4)  & 11(2)   &  
                                    9.10(4) & 11.0(3)& 12.0(6)\\[0.1in]
$r_1$ \dotfill &                     \nodata & 0.55(10)& 0.42(8) & 
                                     \nodata & 0.258(16) & 0.28(2) \\
$\lambda_1$ (\AA) \dotfill &         \nodata & 27.3(3) & 25.71(15) & 
                                     \nodata & 24.75(15)\tablenotemark{a} & 25.4(2)  \\
${\rm FWHM}_1$ (\AA) \dotfill&       \nodata & 17(3)   & 13(2) &     
                                     \nodata & 6.7(5)\tablenotemark{a}  & 8.9(12)  \\
${\rm EW}_1$ (\AA) \dotfill &        \nodata & 10      & 6 &         
                                     \nodata & 1.8     & 2.7\\[0.1in]
$r_2$ \dotfill &                     \nodata & \nodata  & 0.27(7) &  
                                     \nodata & \nodata  & \nodata\\
$\lambda_2$ (\AA) \dotfill &         \nodata & \nodata  & 21.52(6) & 
                                     \nodata & \nodata  & \nodata\\
${\rm FWHM}_{2}$ (\AA) \dotfill &    \nodata & \nodata  & 0.46(16) & 
                                     \nodata & \nodata  & \nodata\\
${\rm EW_{2}}$ (\AA) \dotfill &      \nodata & \nodata  & 0.13 &     
                                     \nodata & \nodata  & \nodata\\[0.1in]
$\chi^{2}$ \dotfill &                445     & 152      & 139 &     
                                     542     & 291      & 281\\
DOF \dotfill &                       170     & 167      & 164 &     
                                     144     & 141      & 141\\
$\chi^{2}/$DOF \dotfill &            2.62    & 0.91     & 0.85 &    
                                     3.77    & 2.06     & 2.00\\
\enddata
\tablecomments{The second row in the header refers to the type of the
  fit: 0-G: black-body continuum only; 
  1-G: including one Gaussian absorption component (in wavelength
  units); 2-G: including two Gaussian absorption components; 1-G$_E$:
  one Gaussian in energy units.  The parameters $r$, $\lambda$,
  FWHM, and EW are the fractional depth, central wavelength, full width
  at half maximum, and equivalent width of the Gaussian feature (for a
  Gaussian, ${\rm EW} = r \sqrt{\pi/4\log2}\,{\rm FWHM}$).  The
  numbers in parentheses are 1-$\sigma$ uncertainties in the last
  digit (determined by varying the parameter while leaving all other
  parameters free).}
\tablenotetext{a}{The Gaussian was written in terms of energies, and
  the best-fit parameters were $E_1=0.493(3)\,$keV, and ${\rm
  FWHM}=0.139(11)\,$keV.  These values were converted to wavelengths
  in the table.} 
\end{deluxetable}

\subsection{Determining the continuum}\label{sec:pnspec}

The EPIC-pn data cover a larger range in energy and hence might help
to determine the shape of the overall continuum.  We fit the four data
sets taken through the thin filter jointly.  For this purpose, we
rebinned all data sets to have a similar number of counts in each bin
and a bin width of $\sim\!20\,$eV, which is roughly one third of the
spectral resolution.  For the three timing-mode data sets we used the
singles plus doubles at $>\!0.5\,$keV, and for the large-window mode
data set we used singles only at $>\!0.15\,$keV.  As expected from the
RGS data, we find that a simple absorbed blackbody does not work: the
implied column density is again $N_{\rm H}=0$, and the reduced
$\chi^2_{\rm red}$ is an unacceptable 4.4 (\Tref{fits}).  The
residuals, shown in \Fref{epic}, deviate significantly from a smooth
spectrum, with the largest deviation at energies near 0.45\,keV
(28\,\AA), just where the RGS data showed the broad absorption feature
(\Fref{rgsmodel}).  In this region, there are no known instrumental
edges, etc.

\begin{figure}
\plotone{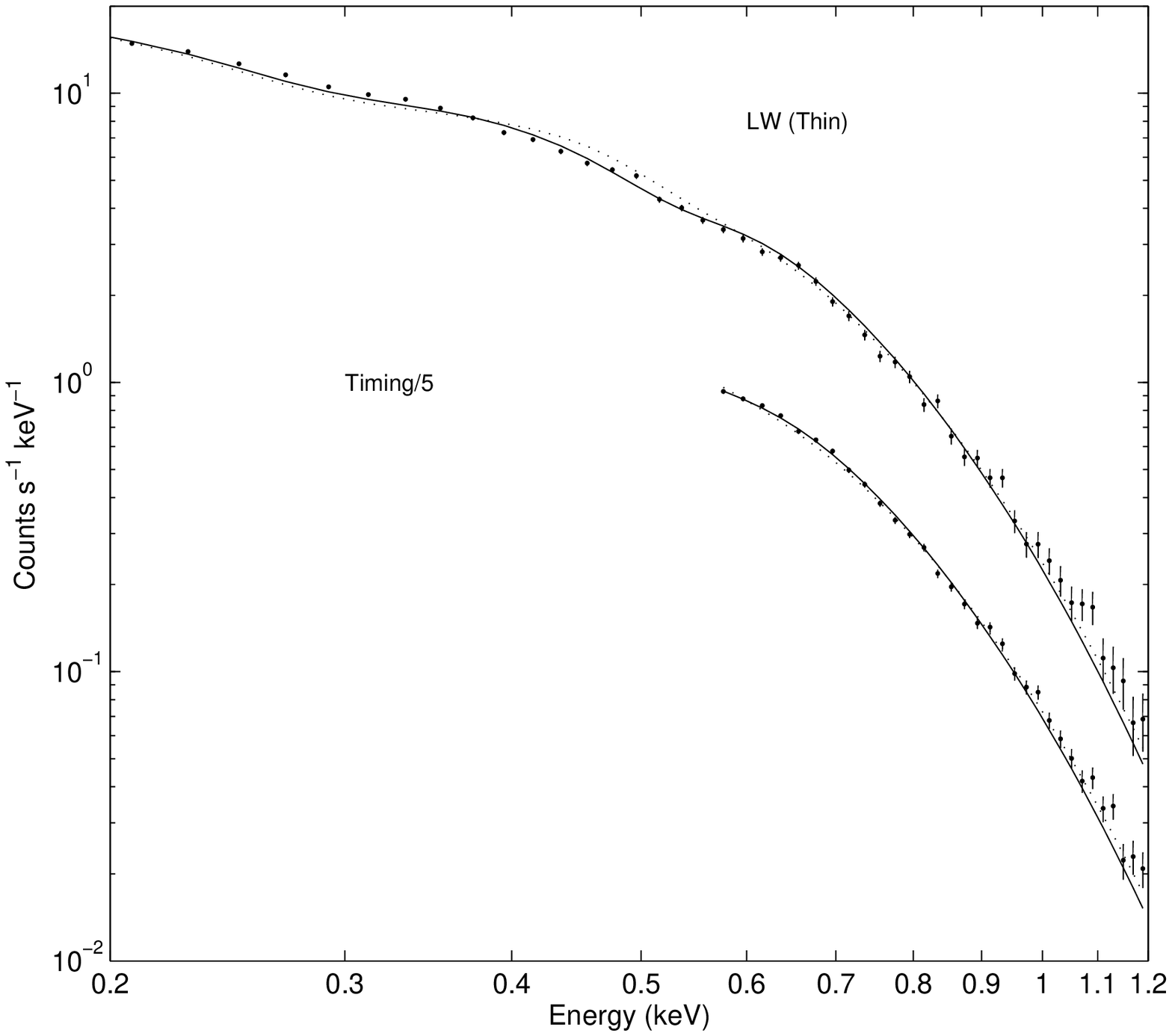}
\plotone{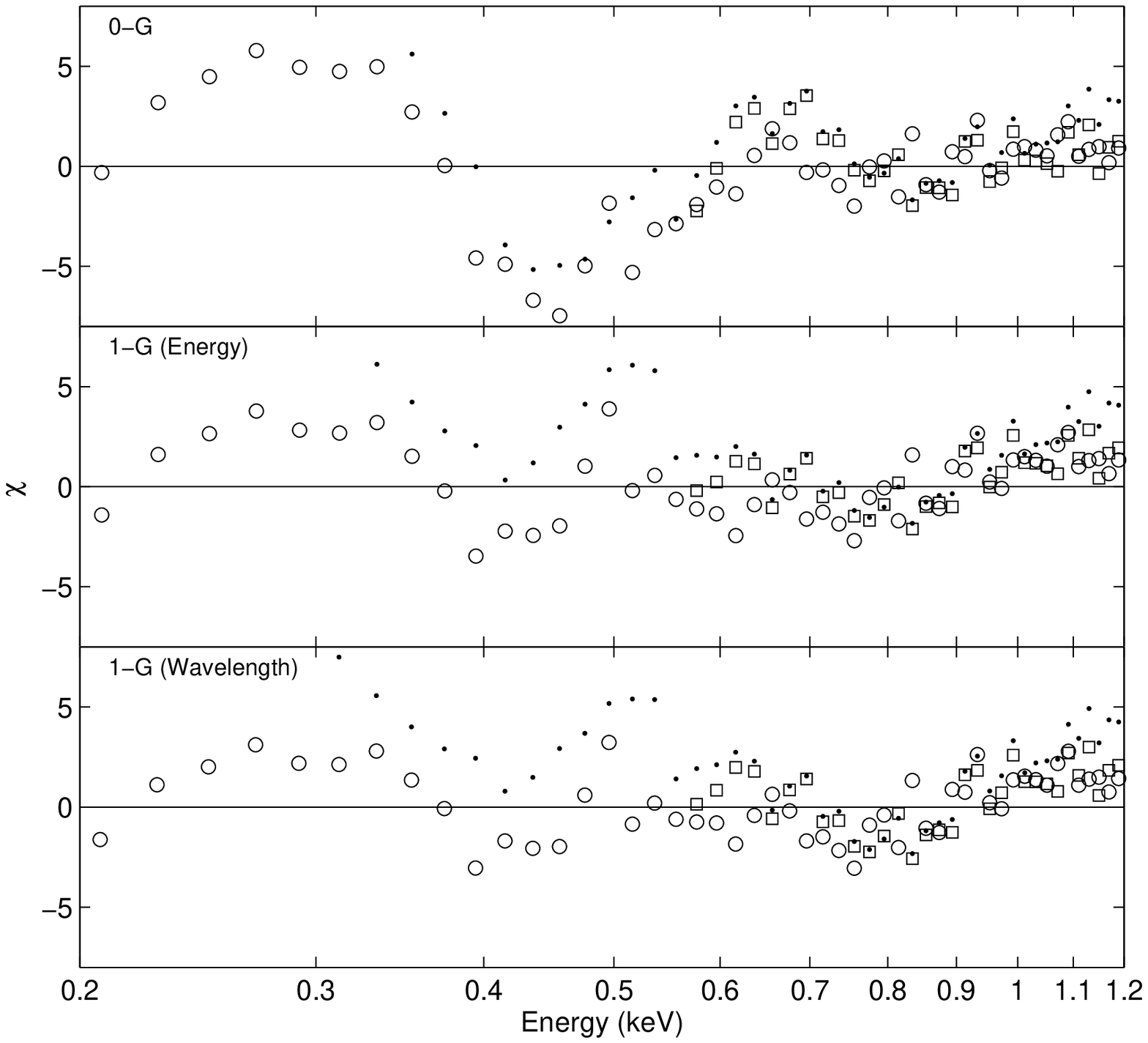}
\caption[]{Fits and residuals to the {\em XMM} EPIC data taken through
the thin filter.  The top panel shows the observations, as well as the
best fits using a single black body (dotted curve) and a black body
with one Gaussian-shaped absorption feature (continuous curve).  The
bottom panels show the residuals for these two cases, as well as for
the best fit of a black body with a single Gaussian in wavelength
units.  The residuals for the observations taken in large-window mode
are shown by circles, those in timing mode by squares.  The residuals
for singles down to 0.3\,keV in timing mode are shown by dots (see
\Sref{spectrum:verification}).\label{fig:epic}}
\end{figure}

As a first-order parametrisation, we again fit a blackbody absorbed by
a Gaussian, described as in \Eref{gauss}, but written in energy units
(i.e., replacing all wavelengths in \Eref{gauss} with energies; the
rationale is that for the EPIC CCD spectra the resolution in energy
$\Delta E$ is roughly constant at low energies, while the resolution
in wavelength $\Delta\lambda$ changes rapidly; for the grating
spectra, the reverse holds).  We find that this leads to a significant
improvement, but that the fit is still unacceptable ($\chi^2_{\rm
red}=2.0$).  Indeed, although the fit to the data looks fairly good
(\Fref{epic}), the residuals show clear systematic deviations.  As the
RGS fit was so much better, we wondered whether it would help to use a
Gaussian in wavelength units rather than energy units (for a wide
Gaussian, the shape is significantly different).  Indeed, for the RGS
data we find that with a Gaussian in energy units, the fit is not very
good (worse than the fit using the Gaussian in wavelength units by
$\Delta\chi^2=10$; best-fit $N_{\rm H}$ remaining at zero).  For the
EPIC data, the fit improves as well ($\Delta\chi^2=26$, no change in
degrees of freedom), but the difference in the results is small (see
\Fref{epic}; \Tref{fits}).  We tried fitting more Gaussians, as well
as different shapes (Lorentzians), but did not find a simple,
significantly better result.

\begin{figure}
\plotone{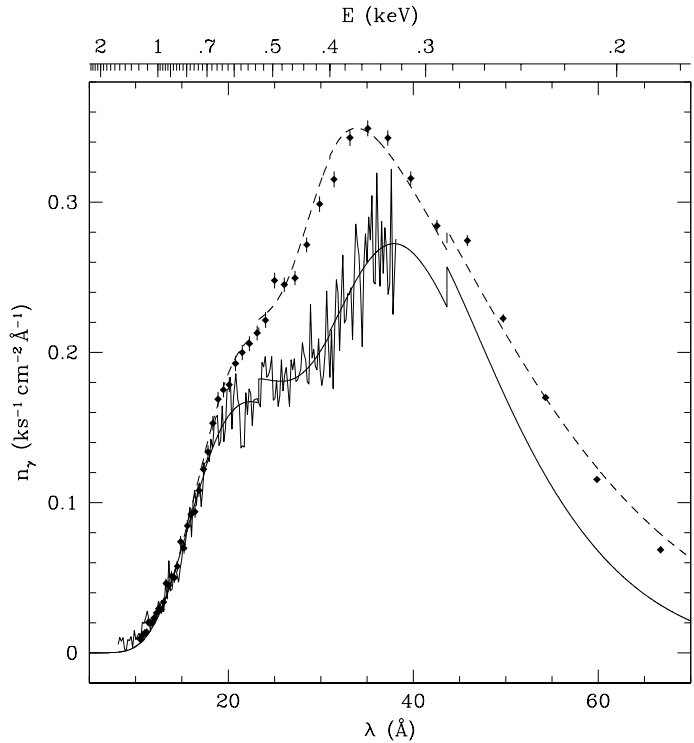}
\caption[]{Unfolded spectra.  The points reflect the EPIC PN data
  obtained in large-window mode, and the dashed curve is the best fit
  to them by an absorbed black body with a single Gaussian absorption
  feature (in wavelength units; \Tref{fits}).  (Note that the
  uncertainties for the unfolded EPIC data are not reliable for the
  large bin sizes at long wavelengths.  The quality of the fit should
  be judged from \Fref{epic}.)  The continuous curves are the
  flux-calibrated RGS spectrum and its best fit (single Gaussian;
  \Tref{fits}).\label{fig:unfolded}}
\end{figure}

What did become clear, however, is that the inferred shape of the
continuum is rather sensitive to the choice of parametrisation.  For
instance, from \Tref{fits}, one sees that the results for $N_H$, $kT$,
and $R/d$ differ significantly depending on whether one uses the
Gaussian in energy or wavelength units.  The difference between the
results from RGS and EPIC, however, is larger still.  Mostly, this
reflects the inconsistencies in the calibration of the two instruments
at low energies mentioned at the start of this section, as can be seen
in \Fref{unfolded}, where we show the unfolded EPIC spectrum taken in
large-window mode, with the flux-calibrated RGS spectrum overdrawn.

Less clear is whether these inconsistencies could explain the
difference in strength of the absorption feature, the equivalent width
for the RGS being a factor three larger than that for EPIC.  In order
to determine the significance, we fitted the RGS data forcing the
equivalent width at the value found by EPIC (3.3\,\AA).  We found that
$\chi^2$ increased by 14, which is not a highly significant change.
However, $N_{\rm H}$ became zero, which is not realistic.  If we also
fix $N_{\rm H}$ to the EPIC value ($1.3\times10^{20}{\rm\,cm^{-2}}$),
the fit does become significantly worse ($\Delta\chi^2=36$).  We
conclude that while we can be confident about the presence and the
central wavelength of the absorption feature, we cannot measure its
strength securely.  In the absence of improvements in the calibration,
observations at high resolution but covering a larger wavelength
range, such as could be provided with the LETG on {\em Chandra}, are
required to resolve this issue.

\begin{figure}
\plotone{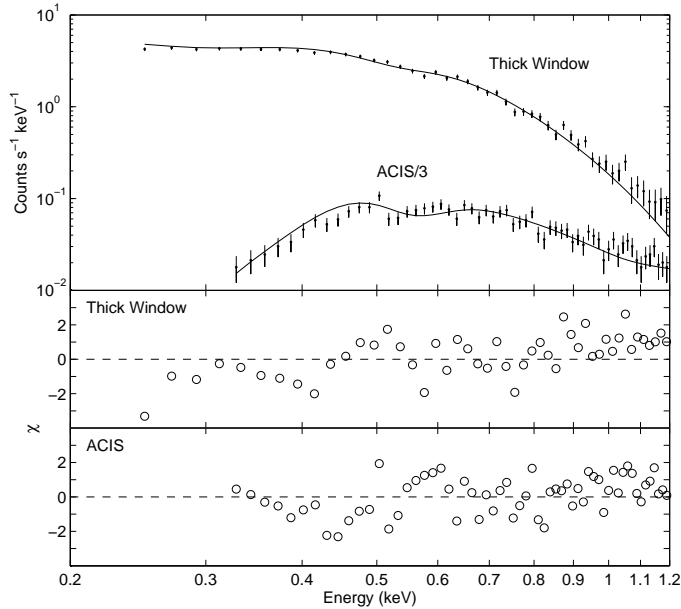}
\caption[]{Comparison of the data taken with {\em Chandra} ACIS-I and
{\em XMM} EPIC through the thick filter with the best fit inferred
from the EPIC data taken through the thin filter (\Fref{epic}).  Both
data sets confirm that a strong absorption feature is present near
0.4\,keV.\label{fig:confirmation}}
\end{figure}

\subsection{Verification}\label{sec:spectrum:verification}

The calibration of the timing mode data, and of observations taken
through the thin filter in general, is still uncertain.  Therefore, we
verified that the black body with a single Gaussian (in energy units)
could reproduce other data sets.  The first is the {\em Chandra} ACIS
data.  For these, there is another free parameter $\alpha$ (the pileup
probability; see \citealt{dav01}).  Applying the model to the ACIS
data -- binned to $\sim\!15\,$eV -- while fitting only for $\alpha$,
we achieved a good fit (\Fref{confirmation}): $\chi^{2}=76$ for 70
degrees-of-freedom, with $\alpha=0.76\pm0.05$, which is within the
accepted range of 0.2--0.8.  The uncertainties of the pileup model,
especially around the instrumental features near 0.5--0.7~keV, make
this fit less reliable than the EPIC data, but the fact that the two
agree and that the absorption feature at 0.45~keV is easily seen gives
confidence that the feature is real and that the continuum shape is
close to correct.

Second, we compared the model to large-window data taken with the
thick filter (binned to $\sim\!20\,$eV).  Again, keeping all
parameters fixed, we find $\chi^{2}=74$ for 53 degrees-of-freedom
(\Fref{confirmation}); this is no worse a fit than what we found for
the data taken through the thin filter.

Third, we refitted the timing data with the same model, but now
selecting only singles, at energies above 0.3\,keV.  As mentioned in
\Sref{obs:xmm:pn}, the response for this setup is not well understood.
Nevertheless, the fit is no worse than that for the other data sets,
and the 0.45\,keV absorption feature is again obvious (see
\Fref{epic}).

\subsection{Limit on high energy flux}

Unlike radio pulsars, the X-ray spectra of the thermally emitting
neutron stars do not appear to require any contribution of non-thermal
emission.  This is the case also for \rxj: at energies above 2\,keV,
we can only set limits to the flux.  Our most stringent constraints
come from the EPIC-PN observations in large-window mode.  In the
2--4.5 and 4.5--7.5\,keV ranges, the upper limits on the count rate
are 4 and $5\times10^{-4}{\rm\,s^{-1}}$ (2$\sigma$), respectively,
corresponding to limits on the flux of 3 and
$9\times10^{-15}{\rm\,erg\,s^{-1}\,cm^{-2}}$, respectively (using the
standard {\em XMM} conversion factors, strictly valid only for a
$N(E)\propto E^{-1.7}$ power law absorbed by $N_{\rm H} =
3\times10^{20}{\rm\,cm^{-2}}$).

\begin{figure}
\plotone{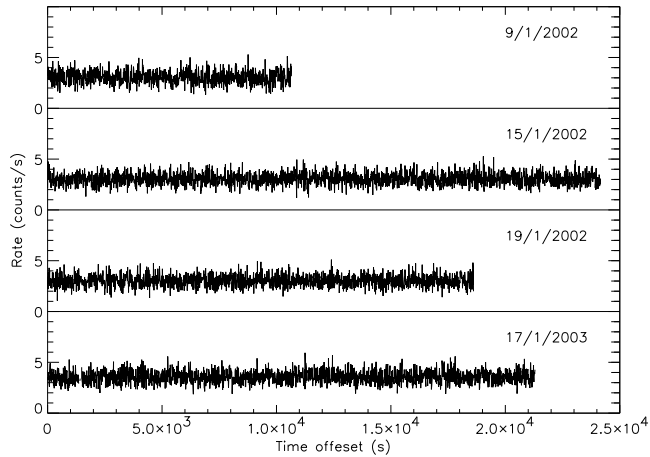}
\caption[]{Lightcurves for each exposure, with net count rates
measured in 10-s bins.  Time is measured from the start of each
observation, and sections near the end that were affected by
background soft proton flares have been removed.  In addition, for the
2002 observations, an energy cut was used to exclude low-energy events
due to hot, flickering pixels.\label{fig:lightcurves}}
\end{figure}

\section{Timing Analysis}\label{sec:timing}

To search for periodic signals, we used the timing-mode data sets, as
well as the data set taken in large-window mode through the thin
filter.  The data set taken through the thick filter had, relatively,
too few counts to be useful.  We barycentered all events using the
XMM-SAS task {\sc barycen}.  For the timing-mode observations, we
included both single-pixel and double-pixel events, but restricted the
energies to ${\rm PI}>300$ and ${\rm PI}>500$, respectively, in order
to exclude the noise close to the threshold mentioned above
(\S\ref{sec:obs:xmm:pn}; \Fref{sd}).  We extracted source and
background counts from the same regions as used for the spectra.  For
the large-window data, we selected both single and double-pixel events
with energies above 0.2\,keV, from within a circle with radius
$45\farcs1$ around the source position.  To calculate net count rates,
we extracted background events from the whole chip, excluding a circle
with radius $98\farcs4$ around the source, as well as the edges of the
chip.

For each observation, we formed lightcurves and used these to remove
sections of the data with significant background flares.  The
remaining parts, displayed in \Fref{lightcurves}, show that the net
source count rate is constant on long time scales.

\begin{deluxetable}{llllll}
\tablecaption{Summary of Timing Observations\label{tab:timing}}
\tablewidth{0pt}
\tablehead{
\colhead{Date} &  \colhead{Exposure} &
\colhead{Counts} & \colhead{Bin Time} & \multicolumn{2}{c}{PF Limit} \\
 & \colhead{(ks)} & & \colhead{(ms)} & \colhead{50\%}
& \colhead{95\%}}
\startdata
\phn9 Jan.\ 2002&  10.0& 30967& 0.60& 0.047& 0.050\\
15 Jan.\ 2002   &  24.0& 75084& 0.71& 0.030& 0.032\\
19 Jan.\ 2002   &  18.5& 56830& 0.55& 0.034& 0.036\\
17 Jan.\ 2003   &  21.0& 79290& 81&   0.024& 0.027\\
\enddata
\end{deluxetable}

Next, for each observation, we formed fast-sampled lightcurves with
bin times (see \Tref{timing}) chosen such that the total
number of bins was $2^n$, suitable for Fast Fourier Transforms (for
the timing mode data, the maximum frequencies were typically
$\sim\!800$~Hz, i.e., above the highest known neutron-star rotation
frequencies).  Where the resulting computed power spectra showed peaks
with single-trial significance in excess of $15\sigma$, we also
computed a $Z_{1}^{2}$ periodogram at that frequency and at
neighboring frequencies (with closer sampling than in the FFT power
spectrum) to mitigate the effects of sampling and ``scalloping.''  The
highest value of $Z_1^2$ was 19 (for the 15 Jan.\ 2002 observation).
Given the number of trials, this is not a significant detection.
Furthermore, marginal peaks did not recur in the different time
series.  Hence, we conclude that there is no significant signal in the
frequency range from 0.01 to $\sim\!800\,$Hz (the precise number
depending on the bin time; \Tref{timing}).

To determine pulsed-fraction\footnote{Here we define the pulsed
fraction to be $({\rm mean}-{\min})/{\rm mean}$, where mean and min
are the mean and minimum flux level of the lightcurve, respectively.}
limits from our power spectra, we simulated a number of event lists
with the same count-rate and exposure times as each observation but
with a sinusoidal signal with given pulsed fraction and frequency
inserted.  After binning and Fourier transforming in manners identical
to those used on the data, we determined what Fourier amplitude
resulted from each pulsed fraction.  Using the exponential statistics
of a power spectrum in the absence of signal, we were able to
determine what Fourier amplitudes resulted in detections with 50\% and
95\% confidence.  This then enabled the determination of what pulsed
fraction could be detected with a given confidence.  These values are
listed in \Tref{timing}.

\section{Implications}\label{sec:speculation}

We presented {\em XMM} observations of \rxj.  Below, we briefly
discuss the implications from the discovery of absorption features in
the spectra, the overall spectral energy distribution, and the lack of
pulsations.

\subsection{Absorption features}

The X-ray spectrum is well represented by a black body with superposed
a broad absorption feature centred at $27\,$\AA\ (0.45\,keV), as well
as possibly a narrow one at $21.5\,$\AA\ (0.57\,keV).  Features
similar to our broad one have been seen in only a few other sources,
in RX J1308.6+2127 and RX J0720.4$-$3125, two other thermally emitting
neutron stars (\citealt{hsh+03,hztb03}), and in 1E 1207.4$-$5209, the
central source in the supernova remnant PKS 1209$-$51/52
(\citealt{spzt02,mdlc+02,bclm03}).  For none of these sources, the
published spectra were of sufficient quality to detect a feature
similar to the narrow feature possibly present at 21.5\,\AA\ in \rxj.
No features at all have been observed for two other thermally emitting
neutron stars, RX J1856.5$-$3754 (the brightest;
\citealt{bzn+01,bhn+03,dmd+02,br02}) and RX J0806.4$-$4125
(\citealt{hz02}), or in the thermal components of the X-ray spectra of
the Vela pulsar (\citealt{pzs+01}) and PSR B0656+14
(\citealt{ms02,pzs02}).

Comparing the energies of the broad absorption features in the three
sources, one finds all are different: $\sim\!0.7$, 1.4, and possibly
2.1 and 2.8\,keV in 1E 1207.4$-$5209, $\la\!0.3\,$keV in RX
J1308.6+2127, $\sim\!0.27\,$keV in RX J0720.4$-$3125, and
$\sim\!0.45\,$keV in \rxj.  Such differences might arise from
differences in magnetic field strength, composition, temperature, or
gravitational redshift.  Given the homogeneity in observed
neutron-star masses (\citealt{tc99}), the last option seems unlikely.
Temperature alone is also unlikely, as the temperatures for \rxj\
(96\,eV), RX J1308.6+2127 (86\,eV), and RX J0720.4$-$3125 (84\,eV) are
similar.  Varying composition is more difficult to exclude, but this
is perhaps rather ad-hoc.  The simplest solution would seem to be
differences in magnetic field, as magnetic field strengths are known
to vary widely.  Indeed, the absence of features in the other sources
mentioned above could be due to the magnetic field being outside of
the range leading to features are in the X-ray band.  Finally, it
might account for the harmonic relation between the different features
in 1E 1207.4$-$5209 (\citealt{bclm03}, but see \citealt{spzt02}).

\subsubsection{\rxj\ in isolation}\label{sec:inisolation}

Comparing our spectra of \rxj\ to model atmospheres, we find that they
are inconsistent with all of the non-magnetic models presented so far
(\citealt{zp02}, and references therein).
For the Hydrogen and Helium models, no features are expected in the
X-ray band, while for the heavy-element compositions that have been
considered, the energies are not right.

For neutron stars with pulsar-like magnetic fields
($\sim\!10^{12}\,$G), models have been calculated for pure Hydrogen
(the more detailed of which take account of the different bound-bound
transitions, etc.; \citealt{zp02}, and references therein), and first
attempts have been made for Iron (\citealt{rrm97}).  For even
stronger, magnetar-strength fields ($B\ga10^{14}\,$G), most models so
far assumed completely ionised Hydrogen (for a review, see
\citealt{zp02}; for first results including neutral Hydrogen, see
\citealt{hlpc03}).  For elements other than Hydrogen, \citet{mh02}
presented detailed calculations of energy levels and transition
probabilities, but these have not yet been used in model-atmosphere
calculations.  Below, we will limit ourselves to a composition of pure
Hydrogen, as this seems the most likely one, given that gravitational
settling will rapidly make the lightest element float to the surface.

\paragraph{Cyclotron absorption.}  
For \rxj, if we make the usual assumption of cyclotron absorption, the
0.45\,keV absorption feature could in principle be due to an electron
cyclotron line in a $\sim\!4\times10^{10}(1+z)\,$G field or a proton
cyclotron line in a $\sim\!7\times10^{13}(1+z)\,$G field (here,
$1+z=(1-2GM/Rc^2)^{-1/2}$ is the gravitational redshift factor, equal
to 1.3 for a $10\,$km, $1.4\,M_\odot$ neutron star).  In this case,
however, the observed width of the line ($\sigma_E/E=1/4\ldots1/8$,
with the range set by the uncertainties in the fits) would be
surprisingly small, given that one expects the magnetic field strength
to vary substantially over the surface -- by a factor two for a
centred dipole (and more for an off-center one) -- and the cyclotron
energy is directly proportional to~$B$.  One could appeal to a
relatively small, hot polar cap, over which the field strength would
vary little, but this seems hard to square with the lack of
pulsations.

\paragraph{Neutral hydrogen.}
If the magnetic field is very strong, neutral Hydrogen is strongly
bound and may well be present in significant amounts at the relatively
low temperature of $\sim\!10^6\,$K; indeed, even molecules may be
present (for a review, see \citealt{lai01}).  For neutral Hydrogen,
the observed energies can be reproduced only by transitions from the
tightly bound ground state.  For its binding energy to exceed
$0.45(1+z)\,$keV, the magnetic field has to exceed $10^{14}\,$G (for
which the binding energy is 0.541\,keV, i.e., this would correspond to
$z=0.23$).

In such strong magnetic fields, possible transitions from the ground
state are either to the continuum and weakly bound states, or to
quasi-bound states (G.\ Pavlov, 2003, pers.\ comm.).  Here and below,
we use numerical values found using the approximations of
\citet{pot98}; for a review of the different types of excitation
states of neutral Hydrogen in a strong magnetic field, see
\citet{lai01}.  We list the transitions to the continuum and weakly
bound states together, since the weakly bound states have binding
energies of order 1\,Ryd and hence the transition energies to those
are similar to the energy of the bound-free transition.  For these
transitions, the width of the feature will depend on two main effects.
First, the binding energy of the ground state varies as $(\log
B/B_0)^2$ (where $B_0=2.35\times10^9\,$G; \citealt{lai01}), and hence
a factor $\sim\!2$ variation in field strength over the surface should
broaden the line by $\sigma_E/E\sim 1/7$.  Second, the binding energy
changes due to the so-called motional Stark effect: the faster an atom
moves, the less bound it is (\citealt{pm93,pp97}); because of this,
the feature is broadened towards lower energies by $\sim\!kT$, i.e.,
$\sigma_E/E\sim 1/5$.  Such a broadening is consistent with that
observed.

Looking in detail at the opacities (\citealt{pp95,pp97}), one might
expect that there would be a narrower component at the blue side,
reflecting the bound-bound transitions from `centered states,' states
little affected by the motional Stark effect.  If we identify these
transitions with the possible 0.57\,keV feature, the implied magnetic
field strength would be about $4\times10^{14}\,$G.  In this case, the
proton cyclotron line would be at 2.5\,keV, i.e., out of the observed
band.  

We caution, however, that in models for pulsar-like field strengths of
a few $10^{12}\,$G, the spectra do not show strong features at the
transitions to weakly bound states or the continuum: the oscillator
strengths are large, but the absorption takes place in parts of the
atmosphere where the temperature gradient is shallow (G.\ Pavlov,
2002, personal comm.; see Fig.~10 in \citealt{zp02}).  In these
models, the transitions to other tightly bound states lead to much
stronger absorption, even though the oscillator strengths are smaller.
The same appears to hold for stronger magnetic fields
(\citealt{hlpc03}).

For field strengths above $\sim\!6\times10^{13}\,$G, the only stable
tightly bound state is the ground state; the higher states are
quasi-bound, auto-ionizing states, as they have energy levels above
the continuum for the ground state.  For the transition to the lowest
quasi-bound state to have energy $\sim\!0.45(1+z)$\,keV, the magnetic
field strength should be $B\simeq7\times10^{13}\,$G.  For such a field
strength, however, the proton cyclotron line would be at
$\sim\!0.3\,$keV and should be noticeable too.  Perhaps more
interestingly, if we identify the 0.45\,keV feature with the proton
cyclotron line in a $B\simeq9\times10^{13}\,$G field (for
$1+z\simeq1.3$), the transition to the lower quasi-bound level would be
at 0.53\,keV, consistent with our marginally detected line.

\paragraph{Vacuum polarisation.}
For both the identification with the proton cyclotron line and with
the features from neutral Hydrogen, the required field strengths are
above the critical field $B_{\rm QED}=4.4\times10^{13}\,$G, at which
the electron cyclotron energy equals the electron rest mass.  In
strong fields, photons propagating down the density gradient in a
neutron star atmosphere can change from one polarization mode to
another at ``vacuum resonance,'' where the plasma contribution to the
dielectric properties is compensated by that due to the quantum
electrodynamics effect of vacuum polarization.  When this resonance
occurs between the deeper photosphere for the extraordinary mode
photons and the shallower photosphere for the ordinary mode ones, it
will reduce the contrast of spectral features (\citealt{hl03}).  For
the energies corresponding to our feature ($\la\!0.7\,$keV at the
surface), this will be important for magnetic fields in the range 0.7
to $50\times10^{14}$\,G (\citealt{lh03}).

Given the above, one would expect putative features due to neutral
hydrogen in a $\ga\!10^{14}\,$G field to have reduced strength.  For
the proton cyclotron line, however, the situation is less clear, as
the inferred magnetic field is close to the lower boundary.  Indeed,
the vacuum resonance phenomenon might be the resolution of the
possible problem mentioned above, that the observed width of the
feature is surprisingly narrow, given that the magnetic field strength
is expected to vary by a factor of two over the surface.  It might
simply be that we are seeing only absorption from regions with
relatively low field, $B\simeq9\times10^{13}\,$G, the contrast of the
absorption in regions with higher field being reduced due to the
vacuum resonance.

\subsubsection{Comparison to RX~J1308.6+2127}

\citet{hsh+03} found an absorption feature at $\la\!0.3\,$keV in
RX~J1308.6+2127, another thermally emitting neutron star.  While this
feature is much stronger and wider than the one we found \rxj, we can
see if we can make progress under the assumption that, despite these
differences, the two features have the same origin.

For RX~J1308.6+2127, we have additional information, viz., the slow
spin period, $P=10.3\,$s.  From the temperature, its age should be
about half a million years; assuming magnetic dipole spin-down down
from an initial rapid spin period, its current spin-down rate should
be $\dot{P}\sim P/2t\sim3\times10^{-13}$, which implies a magnetic
field of $3.2\times10^{19}(P\dot{P})^{1/2}\sim6\times10^{13}\,$G.
\citet{hsh+03} note that this is consistent with an interpretation of
the $\la\!0.3\,$keV feature in terms of proton cyclotron absorption in
a field of $\la\!5\times10^{13}(1+z)\,$G.  

If the picture of \citet{lh03} is correct, then for this field
strength, unlike what was the case for \rxj, the proton cyclotron
absorption in RX~J1308.6+2127 would not be affected by vacuum
resonance mode conversion, and hence should be wide and strong.  And
indeed, as mentioned by \citet{hsh+03}, the feature's width,
$\sigma_E/E\ga1/3$, is consistent with the expected variation of the
cyclotron energy with a factor two over the surface, and the
equivalent width is consistent with model calculations of
\citet{ztst01} (which do not take into account vacuum resonance
effects).  \citeauthor{hsh+03} also mention that the feature extends
up to $\sim\!0.5\,$keV, which is similar to the maximum energy at
which we observe absorption in \rxj.  Thus, it might be that in both
sources the maximum energy out to which absorption is seen is set by
vacuum resonance, but that in RX~J1308.6+2127 the part of the surface
with higher proton cyclotron energy is small, while for \rxj\ it is
large, thus leading to a strong and wide absorption feature in the
former source, and a weak and narrow one in the latter.

We now consider the case of neutral Hydrogen.  As lines and edges are
expected to be broadened redward, the relevant energy is the highest
one at which absorption is observed, i.e., $\sim\!0.5\,$keV for
RX~J1308.6+2127 (\citealt{hsh+03}).  This is similar to what is seen
for \rxj, and thus one would infer a similar magnetic field strength,
$\sim\!10^{14}\,$G.  This would seem hard to square with the observed
differences in strength and width of the features.  However, one
should keep in mind that the binding energy is not very sensitive to
$B$, and hence the field-strength estimate is highly uncertain.
Furthermore, the temperature of RX~J1308.6+2127 is slightly lower than
that of \rxj, and hence the neutral fraction may be substantially
higher.  If neutral Hydrogen were indeed responsible for the feature,
X-ray spectra at higher signal-to-noise ratio might reveal a narrow
feature at the blue edge.

\subsubsection{Comparison to RX J0720.4$-$3125}

While revising our manuscript, a number of new results appeared.
First, \citet{hztb03} found an absorption feature in another isolated
neutron star, RX J0720.4$-$3125.  The feature, centered at
$271\pm14$\,eV, with a FWHM of $151\pm16$\,eV and an equivalent width
of $\sim\!40\,$eV, is rather shallow and broad, which may explain why
earlier {\em XMM} (\citealt{pmm+01}) and {\em Chandra}
(\citealt{pzs02,kvkm+03}) observations had failed to detect it.
Second, \citet{dvvmv04} found evidence for long-term changes in the
{\em XMM} RGS spectra as well as the EPIC pulse profile, with the
deviation from a simple black-body spectrum and the pulsed fraction of
the flux increasing in time.  Third, \citet{hl04} presented more
advanced calculations of the effects of vacuum resonance and make
qualitative comparisons with the observations of absorption features
in all three sources.

These new results allow us to test the semi-empirical ideas described
above.  The test is made particularly interesting by the fact that RX
J0720.4$-$3125 has a temperature ($kT=84\,$eV) that is very similar to
what is observed for the other two sources.  We first consider the
simplest idea described above (and discussed also by \citealt{hl04}),
that the features are due to proton cyclotron absorption, reduced in
contrast due to vacuum resonance mode conversion for strong magnetic
fields.  We find an immediate problem: for RX J0720.4$-$3125, the
absorption feature is weaker than in RX J1605.3+3249, yet its central
energy is lower, which would imply a weaker magnetic field and hence
less reduction in contrast due to vacuum resonance.  Indeed, the
central energy is closer to that seen for RX J1308.6+2127, and one
would thus have expected -- all other things being equal -- a
similarly strong absorption feature.  \citet{hl04} suggest that the
feature in RX J0720.4$-$3125 may be weaker because its magnetic field
distribution is non-uniform, with only small patches of the surface
having $B\la10^{14}\,$G, low enough to cause absorption.  This cannot
be excluded, but seems somewhat ad-hoc.  Below, we explore an
alternative explanation.

Before continuing, we note that, like for RX J1308.6+2127, we have
additional information for RX J0720.4$-$3125, viz., its spin period of
8.39\,s.  This period is close to that of RX J1308.6+2127, and since
the temperature -- and therefore age -- are similar as well, one
infers that the magnetic field of RX J0720.4$-$3125 should also be
$\sim\!6\times10^{13}\,$G.  Given the slightly shorter spin period and
slightly lower temperature (84\,eV vs.\ 86\,eV; \citealt{hab04}), the
magnetic field of RX J0720.4$-$3125 should be the weaker of the two.
A weaker field, $B<6\times10^{13}\,$G, is also inferred from pulse
timing (\citet{kkvkm02}).

Given the relatively low inferred field strength, vacuum resonance
should not affect the strength of spectral features.  To avoid seeing
strong proton-cyclotron absorption, therefore, it seems necessary for
the proton-cyclotron line to be outside the observed band, i.e., for
the star to have lower magnetic field.  If so, the absorption could be
due to neutral Hydrogen.  Using the approximations of \citet{pot98},
we find that for $B=4\times10^{13}\,$G the observed energy of 271\,eV
($\sim\!350\,$eV on the surface) could be matched by a transition from
the ground state to the first excited tightly bound state.  For this
field, however, the proton cyclotron line is at $\sim\!0.2\,$keV,
which is still in the observed band.

From the model calculations of \citet{hlpc03}, it seems that the
transition from the ground state to the second excited tightly bound
state can also lead to fairly strong absorption.  This transition is
at the observed energy for $B=1.8\times10^{13}\,$G.  For this field,
the proton cyclotron line is at 90\,eV, i.e., out of the observed
band.  However, one would expect the transition to the first excited
state, at $\sim\!150\,$eV, to be visible.  Given that the low-energy
end of the spectrum is not very well constrained, this may actually be
the case.  For completeness, we note that for this field strength, the
ionisation edge is at $\sim\!275\,$eV, and might contribute to the
observed absorption feature as well.

If neutral Hydrogen is present in sufficient abundance to cause the
absorption in RX J0720.4$-$3125 in a (relatively) weak magnetic field,
it should also be present in the other two sources, since those have
similar temperatures, but, under the present hypothesis, stronger
magnetic fields and hence higher Hydrogen ionisation energies.  For
RX~J1308.6+2127, with $B\simeq6\times10^{13}\,$G, the transition to
the first excited tightly bound state would be at $370\,$eV, i.e.,
together with the proton cyclotron line it would be responsible for
the absorption feature seen.  Time-resolved spectra at higher
resolution and signal-to-noise might separate the two.  The
ionisation edge would be at similar energy, while the transition to
the second excited tightly bound state would be at $\sim\!0.7\,$keV
(this state would be strongly auto-ionising).  As mentioned in
\Sref{inisolation}, for RX J1605.3+3249, with
$B\simeq9\times10^{13}\,$G, the transition to the first excited but
only quasi-bound level would be at 0.53\,keV, which would be
consistent with our marginally detected line.

While the above presents what seems a consistent picture of the
absorption features seen in the three sources, it does not offer an
explanation for the long-term variation in spectral shape and pulse
profile observed in RX J0720.4$-$3125 by \citet{dvvmv04}.  We are very
puzzled by this.

\subsubsection{Other thermally emitting neutron stars}

Continuing the above reasoning to RX J1856.5$-$3754, for which no
features are observed, we conclude that its magnetic fields should be
either $\la\!10^{13}\,$G (to move both proton cyclotron and neutral
Hydrogen features out of the observed band), or $\ga\!10^{14}\,$G (so
that vacuum resonance mode conversion can make the features
unobservable).  Only the former solution is consistent with other
limits: \citet{vkk01b} and \citet{kvka02} use the H$\alpha$ nebula
associated with RX~J1856.5$-$3754a to set constraints on the
energetics, which, combined with an estimate of the age, imply a limit
to the magnetic field strength of
$B\la1\times10^{13}(d/140{\rm\,pc})^{-3/2}\,$G.

The same limits on magnetic field strength might be inferred for RX
J0806.4$-$4123, for which \citet{hz02} failed to find any features.
Looking in detail at their results, however, it seemed that the
residuals to the black-body fit had systematic deviations similar to
those seen in \Fref{epic}.  We wondered whether there might be an
absorption feature after all, at an energy similar to the one in \rxj.
F.\ Haberl (2003, pers.\ comm.) informed us, however, that these
residuals were likely due to calibration uncertainties.  With the
current calibration (which we used as well), the spectrum appears to
be consistent with that of a featureless black body.

\begin{figure}
\plotone{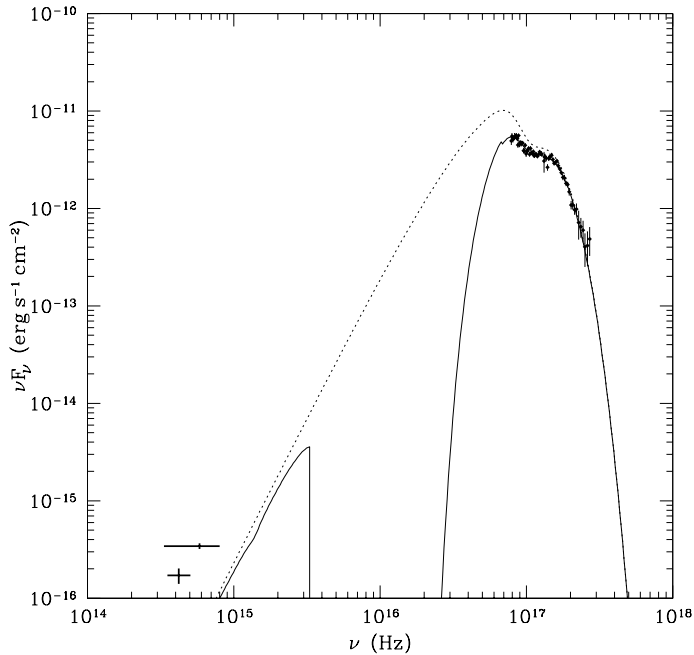}
\caption{Spectral energy distribution.  The points with error bars reflect the
  fluxes inferred from the {\em HST} measurements (\citealt{kkvk03})
  and the RGS measurements (binned to 0.41\,\AA).  The continuous curve is
  the best fit to the RGS data: an absorbed black body with a single
  Gaussian absorption feature (\Tref{fits}).  The dotted curve is the
  same fit, but without interstellar absorption.\label{fig:sed}}
\end{figure}

\subsection{Spectral energy distribution}

The broad-band energy distribution gives additional information about
the nature of the atmosphere.  In \Fref{sed}, we show our RGS
spectrum, as well as the optical/ultra-violet fluxes obtained by
\citet{kkvk03}.  Neither X-ray nor optical provides any evidence for
non-thermal emission.\footnote{We searched for non-thermal radio
emission at 1.4~GHz using the Very Large Array, on 14 February 2001
and 29 April 2001 (in the B and BnA configurations, respectively).
Our cleaned map shows no source at the position of \rxj, with an rms
of $25{\rm\,\mu Jy}$.}  Overdrawn is the best fit to the RGS data,
composed of a black-body with one absorption line (\Tref{fits}).
Apart from the absorption, the spectral energy distribution is similar
to what is observed for other thermally emitting neutron stars
(\citealt{pzs02} and references therein): the X-ray part is well
described by black-body emission, but the optical flux is
underpredicted.

Like the detailed spectrum, the broad-band energy distribution cannot
be matched by any non-magnetic model (\citealt{pzs02} and references
therein).  Hydrogen and helium models, which have hard tails due to
the $\nu^{-3}$ dependence of free-free opacity, overpredict the
optical emission.  If a strong magnetic field is present, at low
energies the opacity is reduced by a factor proportional to
$(\nu/\nu_{\rm cyc})^{-2}$ (where $\nu_{\rm cyc}$ is the electron
cyclotron frequency), and hence the discrepancy can be reduced.
Indeed, for suitably chosen, strong fields ($\ga\!10^{14}\,$G), models
qualitatively matching the X-ray spectrum can also reproduce the
optical (W.~Ho, 2002, personal comm.).  In detail, however, these
models fail at high energies, where the predicted flux drops somewhat
less fast than the observed very steep decline.

One might appeal to vacuum resonance again, as this also affects the
continuum, making the spectrum softer (\citealt{hl03}).  If so,
however, one would expect the high-energy tail in RX J1308.6+2127 (and
perhaps in most others) not be similarly affected, and thus be harder.
This, however, is not the case: all sources have spectra well
described by Wien tails at high energies.  This remains a puzzle.

\subsection{Lack of pulsations}

We did not detect any pulsations, setting an upper limit to the pulsed
fraction of $3\%$ in the frequency range 0.01--800\,Hz.  The same is
true for RX J1856.5$-$3754, where \citet{bhn+03} set an even more
stringent limit of 1.3\% to the pulsed fraction in the range
0.001--50\,Hz, but unlike what is seen for RX J0720.4$-$3125 (8.39\,s,
10\%; \citealt{hmb+97}), RX J0806.4$-$4123 (11.4\,s, 6\%;
\citealt{hz02}), RX J1308.6+2127 (10.3\,s, 20\%; \citealt{hsh+03}),
and RX J0420.0$-$5022 (3.45\,s, 12\%; \citealt{hab04}).

The non-detection of pulsations in RX J1856.5$-$3754 has been a
surprise, given the relatively large pulsed fractions observed for the
other thermally emitting neutron stars, as well as, indeed, for almost
all isolated neutron stars studied in sufficient detail.  Given that
the spectrum seems to require a strong magnetic field, one expects a
temperature distribution over the surface, with the magnetic poles
likely hotter due to decreased opacities in the photosphere, increased
conduction beneath it, and possibly increased heating from the
magnetosphere.  Thus, pulsations are expected.

For lack of alternatives, the absence of pulsations has been
attributed to unfavorable geometry, i.e., a close alignment of the
rotation axis either with the magnetic axis, or with the line of sight
(\citealt{rgs02,br02}).  According to \citet{bhn+03}, the {\em a
priori} probability of obtaining a pulsed fraction as low as observed
is only 1\%.  With our non-detection of pulsations in \rxj, this
explanation has become unlikely.

Could it be that these two objects are rotating extremely slowly?
This would be odd, though not unprecedented: while most white dwarfs
rotate with periods of about one day, a subset of strongly magnetized
white dwarfs hardly rotates at all, with lower limits to the periods
of a century (\citealt{wf00} and references therein).  Perhaps the
same holds for neutron stars.  Or perhaps, as suggested by
\citet{mr03}, the neutron star had such a strong magnetic field that
it could be stopped after its formation, braking on the interstellar
medium.

Alternatively, could it be more likely than thought that one sees no
modulation even though there are temperature differences?  The usual
assumption is that there are two hot polar caps in the midst of a
surface of otherwise uniform temperature.  It has been shown in a
number of studies that the observed pulsed fraction depends strongly
on gravitional bending, with the details depending on the anisotropy
of the emission (e.g., \citealt{pfc83,zsp95}; \citealt{bel02} for
analytic approximations).  Indeed, for a range of inclinations and
magnetic latitudes, gravitational bending ensures the sum of the
effective areas of the two hot regions remains constant, which, if the
emission is isotropic, implies zero modulation.  For instance, for a
radius equal to three Schwarzschild radii (12.4\,km for a
$1.4\,M_\odot$ neutron star), in 25\% of the phase space one would not
observe any pulsations (\citealt{bel02}).  The reason for this
probability being much larger than the one quoted above for
RX~J1856.5$-$3754, is that \citet{br02} and \citet{bhn+03} assume a
much larger radius, of $\sim\!15\,$km, based on the requirement that
the optical flux be reproduced by black-body emission from the cooler,
larger area outside of the hot spots.  This requirement seems risky,
however, as long as we do not understand the emission from the neutron
star atmosphere.

Thus, the lack of pulsations in two objects may indicate that neutron
stars are fairly compact.  There is a possible problem, however, in
that the more compact a neutron star is, the smaller its maximum
pulsed fraction.  For the above numbers, the maximum is about 20\%
(\citealt{bel02}).  This was marginally inconsistent with the
$43\pm14$\% observed in {\em ROSAT} observations of RX J0420.0$-$5022
(\citealt{hpm99}), but more recent {\em XMM} observations showed that
the original pulse period identification was wrong; at the correct
period, of 3.45\,s, the pulsed fraction is $\sim\!12\%$
(\citealt{hab04}), which is consistent with the above limit.  This is
encouraging, but one should keep in mind that in other objects, the
fields are not well described by centred dipoles.  For instance, for
magnetic white dwarfs the fields can be modelled well by dipoles
offset by 10--30\% from the centre (\citealt{wf00}).  Furthermore, for
a magnetic atmosphere, the emission may well be far from isotropic
(e.g., \citealt{zsp95}; for a review, \citealt{zp02}).  Phase-resolved
modelling of the spectra, in particular of the absorption features,
may shed light on the precise geometry.

\subsection{Future work}

We have suggested that the absorption features in RX J1308.6+2127, RX
J0720.4$-$3125 and \rxj\ all arise in a pure Hydrogen atmosphere, with
the absorption dominated by neutral Hydrogen transitions in RX
J0720.4$-$3125, and by proton cyclotron absorption in RX J1308.6+2127
and \rxj.  For the latter source, we suggested the feature is weakened
considerably due to the effects of vacuum resonance mode conversion, a
genuine strong-field quantum electrodynamics effect.  These
suggestions could be confirmed by detailed model atmosphere
calculations (for first results, \citealt{hlpc03,hl04}), as well as
further observations.

One could look for spectral features in the few remaining sources.
Perhaps more interestingly, though, would be high signal-to-noise
spectra of the brighter sources with the low energy transmission
grating (LETG) on {\em Chandra}.  This would extend the range for
which one has high resolution to lower energies, and thus give a
better constraint on the shape of the absorption feature (especially
important for RX J1308.6+2127), and allow one to measure the variation
with pulse phase in detail.  Furthermore, they would help to confirm
or refute the possible narrow feature in \rxj, and search for evidence
of absorption by (different) transitions of neutral Hydrogen, or even
Hydrogen molecules.  Hopefully, this would also give us a clue to what
causes the high energy tails of the X-ray spectra to match Wien tails
so well, and what is responsible for the optical excess over the
extrapolated black-body curves.

For all sources, it would be very useful to have independent estimates
of the magnetic field strength.  For this purpose, further timing
studies would be required.  Furthermore, astrometric studies would
help to measure distances and constrain places of origin and ages.
The key step forward, however, would be X-ray polarization
observations, with which the ideas discussed here can be tested
experimentally. 

\acknowledgments We thank Dong Lai, George Pavlov, Kaya Mori, and Wynn
Ho for very useful discussions about neutron-star atmospheres, and the
referee for a helpful report.  We acknowledge the {\em XMM} user
support group for providing details about the EPIC-PN timing mode, and
thank Frank Haberl, Jan-Willem den Herder and Masao Sako for useful
discussions about {\em XMM} calibration issues at low energies.  This
paper is based on observations obtained with {\em XMM-Newton}, an ESA
science mission with instruments and contributions directly funded by
ESA Member States and NASA.  We also used data from the National Radio
Astronomy Observatory, which is a facility of the National Science
Foundation operated under cooperative agreement by Associated
Universities, Inc., as well as archive data from {\em Chandra}.  We
made extensive use of the SIMBAD and ADS data bases.  We acknowledge
support through a guest observer grant from NASA, as well as through
individual grants from NSERC and the University of Toronto (M. H. v.\
K.), from the Fannie and John Hertz Foundation (D. L. K.), and from
NASA and NSF (S. R. K., F. P.).

\bibliographystyle{apj}

\bibliography{ins}

\clearpage

\end{document}